\documentclass[10pt,a4paper,twocolumn,table]{article}

\usepackage{titlesec}
\usepackage{color}
\usepackage[utf8]{inputenc} 
\usepackage[affil-it]{authblk}

\usepackage{lipsum}
\newcommand\blfootnote[1]{%
  \begingroup
  \renewcommand\thefootnote{}\footnote{#1}%
  \addtocounter{footnote}{-1}%
  \endgroup
  }
\usepackage{siunitx}
\usepackage{float}
\usepackage{tikz}
\usepackage{xcolor}
\usepackage{comment}

\usepackage{braket}
\usepackage{nicefrac}
\usepackage[english]{babel} 
\usepackage[top=2.8cm,bottom=2.8cm,left=2.cm,right=2.cm]{geometry}  
\usepackage{amsmath,amsfonts,amssymb}  

\usepackage{caption}
\usepackage{subcaption}
\usepackage{graphicx}  
\graphicspath{{./Figures/}{./PSTricks/}}  
\usepackage{multirow}
\usepackage{appendix}
\usepackage{wrapfig}
\usepackage{setspace}

\usepackage{csquotes}
\usepackage{cite}

\usepackage{array,booktabs}

\usepackage{enumitem}
\usepackage{caption3} 
\DeclareCaptionOption{parskip}[]{}
\usepackage[font=small,labelfont=bf]{caption}
\usepackage[final]{pdfpages}
\usepackage[colorinlistoftodos]{todonotes}
\usepackage{tikz}
\usepackage{sidecap}
\sidecaptionvpos{figure}{t}
\usepackage{abstract}

\definecolor{mycol}{RGB}{255,230,204}

\usepackage{mdframed}

\usepackage[pdftex]{hyperref} 

\title{\textbf{Boosting the secret key rate in a shared quantum and classical fibre communication system}}
\author{Davide Bacco$^*$$^1$$^{\triangleright}$, Beatrice Da Lio$^*$$^1$$^{\triangleright}$, Daniele Cozzolino$^1$, Francesco Da Ros$^1$, Xueshi Guo$^2$, Yunhong Ding$^1$, Yusuke Sasaki$^3$, Kazuhiko Aikawa$^3$,  Shigehito Miki$^4$, Hirotaka Terai$^4$,\\Taro Yamashita$^5$, Jonas S. Neergaard-Nielsen$^2$, Michael Galili$^1$, Karsten Rottwitt$^1$, \\Ulrik L. Andersen$^2$, Toshio Morioka$^1$, Leif K. Oxenløwe$^1$ 
}
\affil{\small $^1$ CoE SPOC, Dep. Photonics Eng., Technical University of Denmark, Kgs. Lyngby 2800, Denmark}
\affil{\small $^2$ CoE bigQ, Dep. of Physics, Technical University of Denmark,  Kgs. Lyngby 2800, Denmark }
\affil{\small $^3$ Advanced Technology Laboratory, Fujikura Ltd., 1440, Mutsuzaki, Sakura, Chiba, 285-8550, Japan}
\affil{\small $^4$ Advanced ICT Res. Inst., National Institute of Information and Communications Technology, 588-2 Iwaoka, Nishi-ku, Kobe, 651-2492, Japan}
\affil{\small $^5$ Department of Electronics, Graduate School of Engineering, Nagoya University, Furo-cho, Chikusa-ku, Nagoya, Japan}

\date{\vspace{-1em} \small   Dated: \today }

\begin{document}
\pagestyle{plain}
\setcounter{page}{1}
\twocolumn[ 
\begin{@twocolumnfalse}
\maketitle
     \vspace{-0.8cm}
  \begin{abstract}
      \normalsize
         \vspace*{-1.0em}
\noindent During the last 20 years, the advance of communication technologies has generated multiple exciting applications. However, classical cryptography, commonly adopted to secure current communication systems, can be jeopardized by the advent of quantum computers. 
Quantum key distribution (QKD) is a promising technology aiming to solve such a security problem. Unfortunately, current implementations of QKD systems show relatively low key rates, demand low channel noise and use ad hoc devices.
In this work, we picture how to overcome the rate limitation by using a 37-core fibre to generate 2.86 Mbit s$^{-1}$ per core that can be space multiplexed into the highest secret key rate of 105.7 Mbit s$^{-1}$ to date.
We also demonstrate, with off-the-shelf equipment, the robustness of the system by co-propagating a classical signal at 370 Gbit s$^{-1}$, paving the way for a shared quantum and classical communication network.
\end{abstract}
  \end{@twocolumnfalse}
 ]

\subsection*{Introduction}
Our\blfootnote{$^*$dabac@fotonik.dtu.dk, bdali@fotonik.dtu.dk \\$^{\triangleright}$ These authors contributed equally to this work} society is based on the continuous exchange of billions of data, and most of them travel in optical fibres. 
Nonetheless, most of the exchanged data are not protected against upcoming threats, \textit{i.e.}, new algorithms able to break current cyphers and the expected availability of quantum computers~\cite{Shor1997}. 
A quantum computer is a machine based on the laws of quantum physics, which will be able to crack some of the current cryptosystems, whose security relies on the limited computational power of an eavesdropper~\cite{kaye2007}.
As derived by Shannon, a way to achieve information theoretical secure communications is to use One-Time-Pad encryption, which requires a pre-shared key of the same size as the message to be sent~\cite{Shannon1949}.
However, other symmetric key algorithms based on computationally hard problems are also used, such as the Advanced Encryption Standard, since they require keys of constant length (\textit{e.g.} 128 or 256 bits)~\cite{daemenAES}.
The symmetric key used by these cryptosystems must be exchanged between the two parties, and this is usually achieved through public-key algorithms, two of the most widely used being the Rivest-Shamir-Adler and the Elliptic Curve Cryptography~\cite{Rivest1978,miller1985ECC,koblitz1987ECC}.
A quantum computer can however break both algorithms, leaving the task of distributing keys to either post-quantum cryptography~\cite{bernstein2017} or to alternative methods such as quantum cryptography.
Within the latter, quantum key distribution (QKD) addresses this challenge by relying on the laws of quantum physics to provide the required information-theoretic security~\cite{Pirandola2019r}. 
During the last 30 years, multiple demonstrations of free-space, underwater and fibre-based QKD systems have shown the feasibility of such a technology~\cite{Frohlich2013,Frohlich2017,Boaron2018,Ren2017,Yin2017,Boucharde_under,Ling2017}.
Nevertheless, multiple factors are limiting the global deployment of QKD systems: the low information rate, the short propagation distance and the compatibility with the existing network infrastructure. Indeed, most of QKD implementations have been realized in low noise environments (\textit{e.g.} dark fibres), attesting to the difficulty of integration with the bright signals used in classical communications~\cite{Qiu2017,Dixon2017,Tang2016,Bunandar2018,Lee2016,Yoshino2013,Stucki2009}.
In this work, we show how to overcome the low rate and compatibility limitations by exploiting a 37-core multicore fibre (MCF) as a technology for quantum communications~\cite{Sasaki2017}. This technology allows for efficient key generation, enabling the highest secret key rate presented to date. Moreover, we co-propagate in all the 37 cores simultaneously a high-speed classical signal, showing that the quantum communication is only weakly perturbed by it, paving the way for a full-fleshed implementation in current communication infrastructures. 

\subsection*{Results}
\subsubsection*{Multicore fibre for quantum communications}
\begin{figure*}[t]
\centering
\includegraphics[width=.9\textwidth]{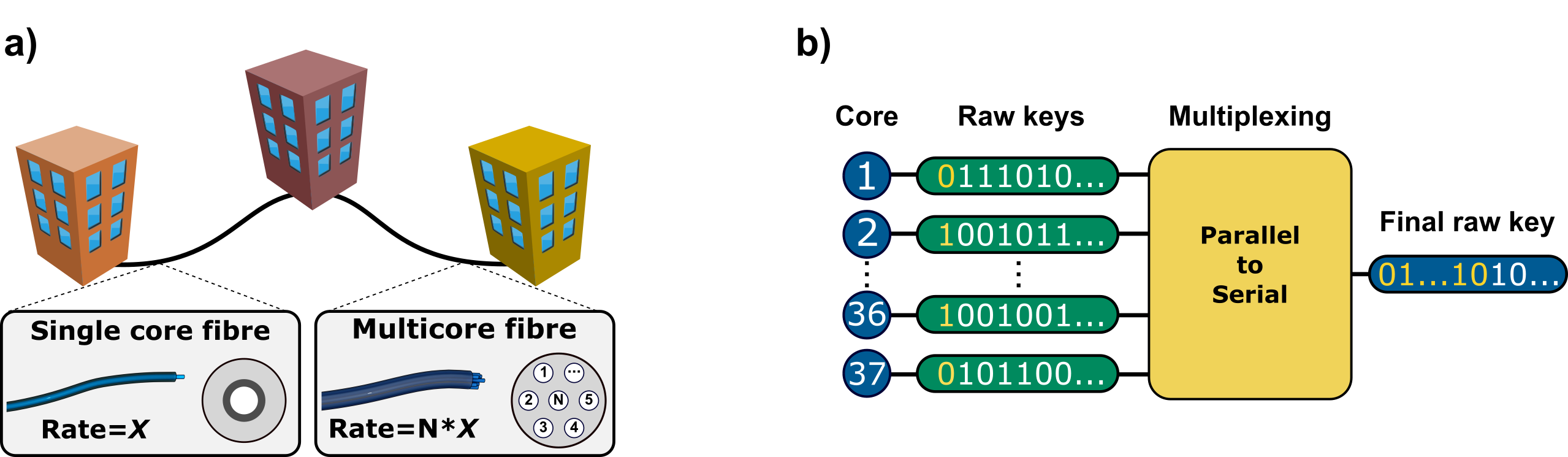}
\caption{\textbf{Multicore fibers applications.} {\bf a) Single mode fibre and multicore fibre.} The fundamental rate-loss law in a single mode fibre is defined by the limit $X= -\log_2 (1-\eta)$, where $\eta$ is the bosonic Gaussian channel transmissivity~\cite{Pirandola2017}. In the case of a multicore fibre (MCF), the total rate is $Rate_{MCF}=\mathrm{N} X$, where N is the number of cores.
{\bf b) Multiplexing of the quantum keys.} Parallel keys are generated in each core of the multicore fibre. After a multiplexing procedure $N$ keys are united in a longer key.}
\label{fig:fig1}
\end{figure*}

\begin{figure}[h]
\centering
\includegraphics[width=0.52\textwidth]{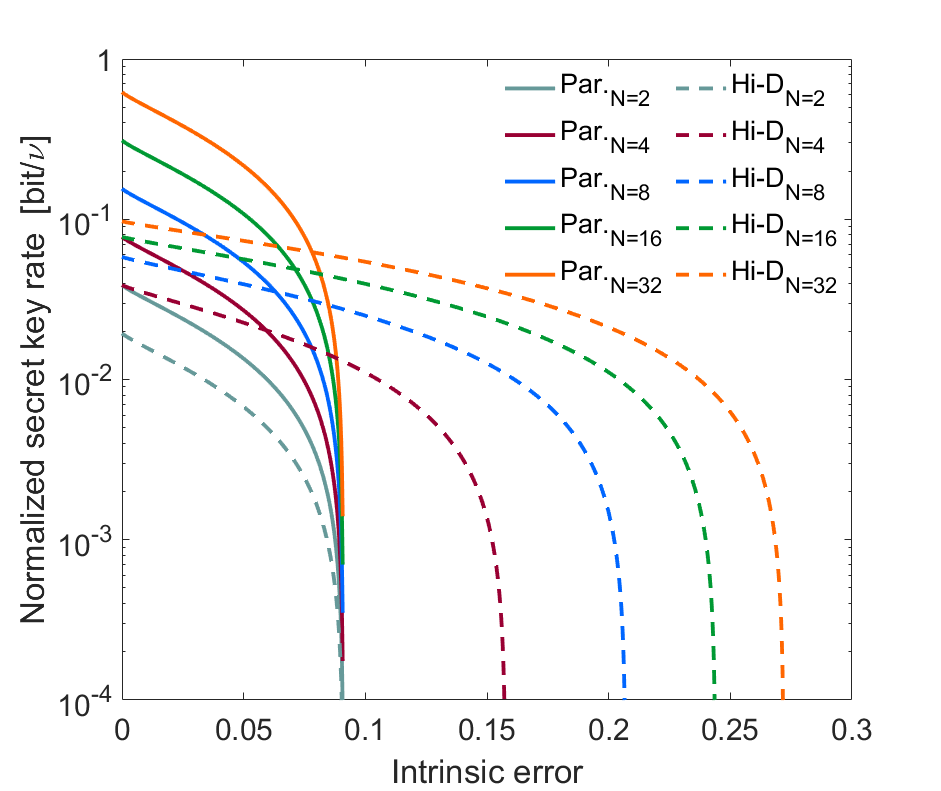}%
\caption{{\bf Simulation of the normalized secret key rate as a function of the intrinsic error for parallel (Par.) and high-dimensional (Hi-D) encoding.} Different colours represent the number of spatial modes used from $N=2$ (light-gray) to $N=32$ (orange). In the Hi-D case (dashed-lines) a multicore or few mode fibres is used to encode the quantum states. A BB84 decoy states protocol with two mutually unbiased bases is implemented for the simulations~\cite{lo2005}. Parameters used are: fibre link $d=7.9$ km (loss $=3.75$ dB), detectors efficiency $\eta_d=0.6$, receiver efficiency $\eta_{bob}=0.85$. In the parallel encoding scheme (full-lines), the same protocol, with same parameters, is used for all the different modes or cores independently. The shown secret key rate is normalized on the system repetition rate $\nu$.}
\label{fig:simulation_rate}
\end{figure}

\begin{figure*}[t]
\centering
\includegraphics[width=0.9\textwidth]{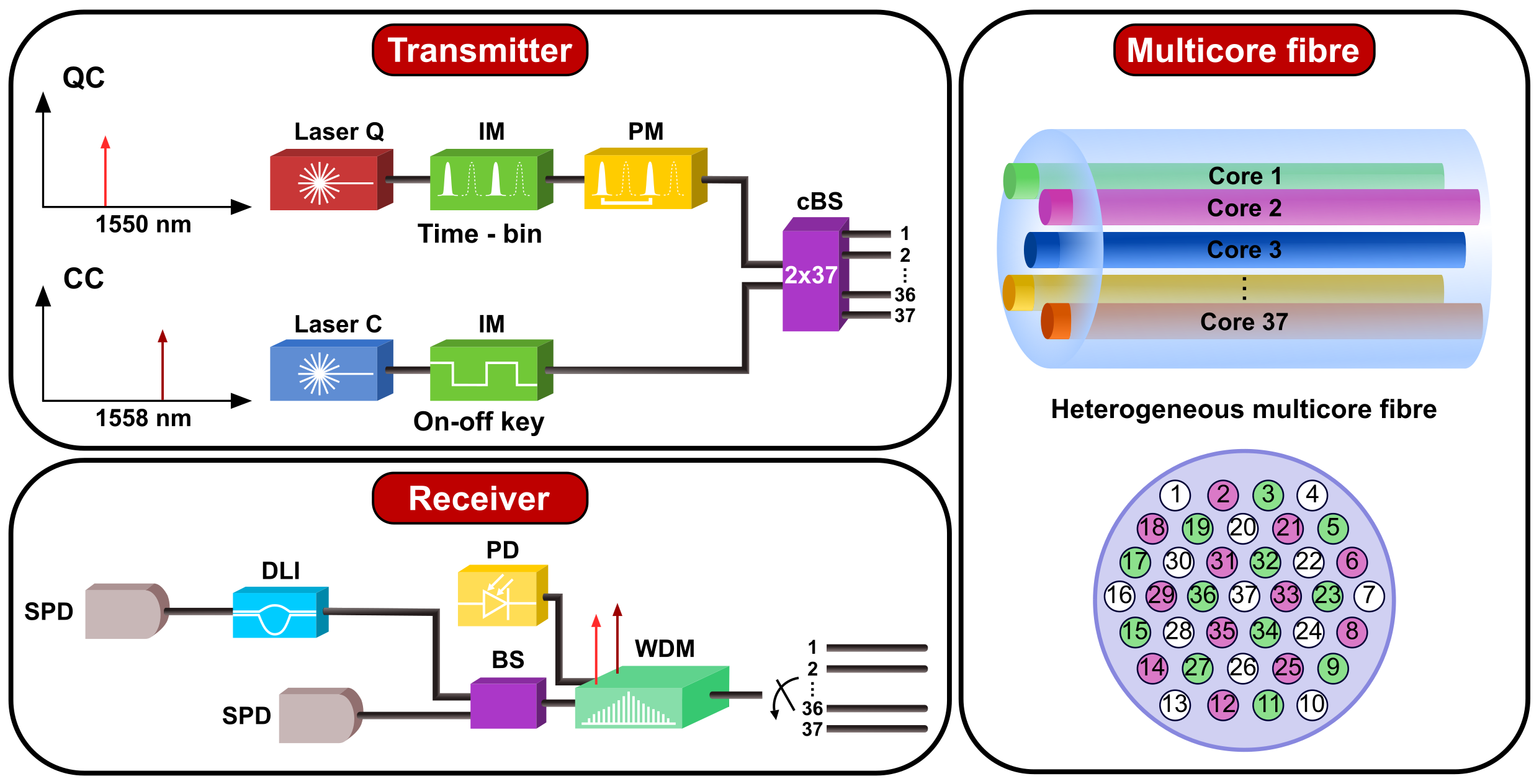}
\caption{{\bf Scheme of the experiment.} A continuous wave laser at 1550 nm (100GHz dense wavelength division multiplexed channel C33) is used for the quantum channel (QC). An intensity modulator (IM) is used for carving the optical pulses and to implement the decoy state method. A phase modulator (PM) is used to phase randomize the pulses. At the same time a different continuous wave laser at 1558 nm (100GHz dense wavelength division multiplexed channel C23) followed by an intensity modulator, controlled by a bit pattern generator, allows a classical channel (CC) with on-off keying modulation at 10 Gbit s$^{-1}$. Through a beam combiner the quantum and classical signals are combined. Cascaded beam splitters (cBS) (two beam splitters 1$\times$5 and four beam splitters 1$\times$8) are used to divide the two signals into 37 cores. The multicore fibre is 7.9 km long and presents slightly different losses between cores. Each core is then received independently and routed to a wavelength division multiplexing (WDM) filter, used to separate the classical and the quantum signals. The classical signal is received by a photo detector (PD) and controlled by an error analyzer. The quantum signal is then measured in two mutually unbiased bases (computational and Fourier bases). A 10 dB beam splitter (BS) is used to divide the two bases: in the computational we only require to measure the time of arrival with a single photon detector (SPD), while in the Fourier basis we use a delay line interferometer (DLI) to detect the phase difference between two pulses (a second SPD is connected to one output of the DLI). All the electrical outputs are collected by a time tagging unit and analyzed by a computer. In the case of the transmission of only the quantum signals, the classical laser was removed in the transmitter and the WDM filter was removed from the receiver.}
\label{fig:setup_simple}
\end{figure*}

\begin{figure}[h]
\centering
\includegraphics[width=0.5\textwidth]{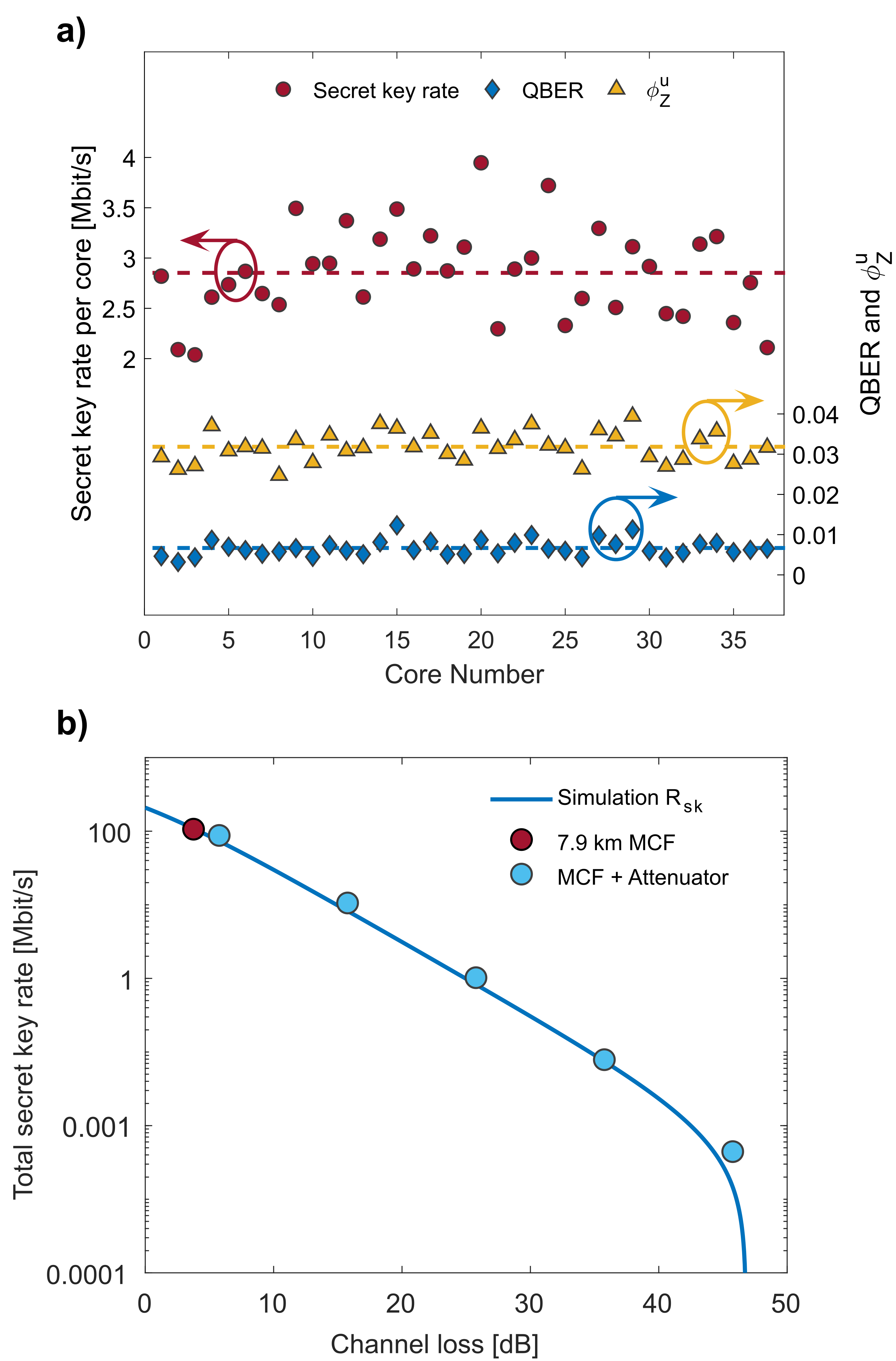}%
\caption{\textbf{Quantum transmission.} {\bf a) Secret key rate (R$_{\text{sk}}$), quantum bit error rate (QBER) and upper bound on the phase error ($\phi_Z^u$) per core.} Red dots represent the secret key rate for each core. The variations are due to different losses of the cores and the slightly different measured values of QBER and visibility. An average value of 2.86 Mbit s$^{-1}$ $\pm$ 4.37 kbit s$^{-1}$ is measured. Blue diamonds are the QBER of the quantum signals with an average value of 0.67\% $\pm$ 0.0038\%, while yellow triangles show the upper bound on the phase error, presenting an average of 3.18\% $\pm$ 0.014\%. Uncertainty values, computed as standard error of the mean, are not displayed as error bars are covered by the markers. {\bf b) Total secret key rate as a function of channel loss.} The red circle is the overall secret key obtained after propagation through the 7.9-km link multicore fibre (MCF): a total of 105.7 Mbit s$^{-1}$ $\pm$ 162 kbit s$^{-1}$ can be achieved through a space division multiplexing scheme over the 37-cores. Light-blue circles are the overall secret key rates expected when the channel is composed by the multicore fibre followed by a variable attenuator to emulate further losses.}
\label{fig:quantum_results}
\end{figure}

Space division multiplexing (SDM) in optical fibre, \textit{i.e.}, communication through distinct and parallel cores/modes in multicore and few-modes fibres, will play a fundamental role in future classical communications, both for solving the future capacity crunch and for reducing the overall power budget, thus decreasing the total power consumption~\cite{Hu2018,Kobayashi2017,Puttnam2015}. As reported in Figure~\ref{fig:fig1} a), through SDM the achievable rate of an N-core MCF corresponds theoretically to N times the achievable rate of a single mode fibre~\cite{Pirandola2017}.
To achieve these high rates, a low cross-talk between the different channels is of extreme importance for SDM schemes. Indeed, cross-talk is the measure of the leakage from one channel to another, leading therefore to a larger insertion loss in the former and, more importantly, to an extra noise source in the latter.
As an example, high-speed classical communication at $10.16$ Pbit s$^{-1}$ has been demonstrated thanks to SDM combined with dense wavelength division multiplexing, polarization modulation and advanced coding ~\cite{Soma2018}.  
Likewise, both multicore and few-mode fibres have been exploited for encoding and transmitting high-dimensional (Hi-D) quantum states~\cite{Ding2016,Cozzolino2018,Canas2016}.In particular, the cores or the modes of these special fibres have been used to increase the dimensionality of the Hilbert space, thus allowing higher photon information efficiency.
Although Hi-D quantum states are suitable for high rate quantum communications, in specific channel conditions (low noise channel), it was experimentally demonstrated that SDM is suitable for very high rate secure communications~\cite{Bacco2017}.
The principle is based on the concatenation of parallel keys acquired in each core/mode as shown in Figure~\ref{fig:fig1} b)~\cite{Bacco2017}. As a support of this statement, Figure~\ref{fig:simulation_rate} shows a theoretical analysis of the secret key rate as a function of the error rate for Hi-D and parallel encoding when assuming transmission in a lossy bosonic Gaussian channel (e.g. a fibre link)~\cite{Pirandola2017}. As a matter of fact, in the low noise regime, the performance achieved by Hi-D protocols is lower compared to the one obtained with the multiplexing of parallel encoding (SDM). Note that, similarly to SDM in classical communications, the generation of parallel keys is achieved when each core/mode is used to transmit independent quantum states. The receiver measures each core/mode individually and then after distillation and reconciliation processes a longer key is available between the two users. In our demonstration we used a 37-core MCF to implement a proof-of-concept experiment where 37 parallel keys, generated in independent cores, are combined.
Table~\ref{tab:comparison} compares this work achievements with state of the art experiments of discrete variable QKD implementations, with or without co-propagation of classical data rate. Details regarding the protocols used, the channel length/loss and the detectors schemes are also included.
As reported in Figure~\ref{fig:setup_simple} a continuous wave laser operating at 1550 nm (100GHz dense wavelength division multiplexing channel C33) followed by an intensity and a phase modulator is used to implement a three-states time-bin protocol, at 595 MHz repetition rate, with 1-decoy state method in finite key regime~\cite{Rusca2018,Rusca2_2018,Boaron2018}. The intensity modulators are used to carve the pulses and to obtain the two (signal and decoy) intensity levels.
We implemented a 1 decoy state protocol, instead of the more common 2 decoy states, because on one side it has been demonstrated that using just 1 decoy does not affect the final system performance (in some cases it even improves them)~\cite{Rusca2018} and on the other side it is easier to implement, in the sense that one intensity modulator driven by a square 2-level signal suffices.
The phase modulator is only used for phase randomization, and it can easily be removed from the setup if a phase-randomized pulsed laser is used as source (i.e. a laser in which the current is switched off between pulses). Cascaded beam splitters allow to divide the signal in 37 different channels and, after being attenuated, these weak signals are injected at the same time each in a different core of the 7.9 km MCF, which has an average loss of 3.75 dB including fan-in/fan-out devices.
This particular MCF presents extremely low cross-talk between different cores thanks to its heterogeneous structure, \textit{i.e.} its particular arrangement of three core types which have different refractive indeces (see Supplementary Note 1 for further details).
At the receiver's side, Bob measures each core independently (one at a time): photons are detected with superconducting nanowire single photon detectors (SNSPDs)~\cite{Miki_2013} and acquired by a time-tagger unit. Note that this experiment represents a proof-of-concept realization. Indeed, in a real implemented system 37 different random sources, transmitters and receivers are required to generate uncorrelated valid keys. More precisely, our implementation would require three intensity modulators (two for pulse carving and one for decoy modulation), one phase modulator and a random source per transmitter, leading to a total of 37 random sources, 111 intensity modulators and 37 phase modulators. Regarding the receiver, 37 unbalanced interferometers and  $(4+1)\times 37=185$ SNSPDs would be needed. Indeed, in the experiment we used four detectors for the first basis and one for the second (see Methods for further details on the experimental setup). On-chip integration of the numerous components seems therefore to be necessary, a solution that is examined in the discussion paragraph.
In Figure~\ref{fig:quantum_results} a) the secret key rate and the quantum bit error rate (QBER) produced per core are reported. Each point is acquired for five minutes of measurement. A total of $105.7$ Mbit s$^{-1}$ $\pm$ 162 kbit s$^{-1}$ is expected to be achieved in the experiment combining all the 37 cores, with a block size of 5.67 Gbit for the finite key analysis (see Supplementary Note 2 for the secret key rate analysis). We chose this specific block size as it corresponds to the overall detection events in the computational basis over a period of 30 seconds, meaning that the acquisition time of our block is exactly 30 seconds. Note that the secret key rate per core shown in Figure~\ref{fig:quantum_results} a) is obtained only from each core raw key through finite key analysis~\cite{Rusca2018,Rusca2_2018,Charles2014}, meaning that error correction and privacy amplification effects are taken into account but not actually implemented and that the obtained key rates also consider fluctuations given by the finite statistics. These rates show an average of 2.86 Mbit s$^{-1}$ $\pm$ 4.37 kbit s$^{-1}$ leading to a total multiplexed rate of 105.7 Mbit s$^{-1}$. The confidence intervals we refer to are derived from the measurements done in the 37 cores of the fiber.
In addition, we use a variable optical attenuator to simulate a longer transmission link and study the scalability of the quantum system. The results are reported in Figure~\ref{fig:quantum_results} b) attesting a channel loss limit of approximately 47 dB ($\approx$ 170 km assuming a MCF average loss coefficient of 0.27 dB/km per core and a constant 1.6 dB of total fan-in/fan-out losses~\cite{Sasaki2017}). In this analysis, the block size is fixed to value of 5.67 Gbit for all points except the last one, thus the block acquisition time increases with the channel losses. To make sure the acquisition time is no longer than 1 day, the block size is reduced to 1 Gbit in the case of the last point (at a channel loss of 45.75 dB).

\subsubsection*{Coexistence of classical and quantum communications}
\begin{figure*}[t]
\centering
\includegraphics[width=\textwidth]{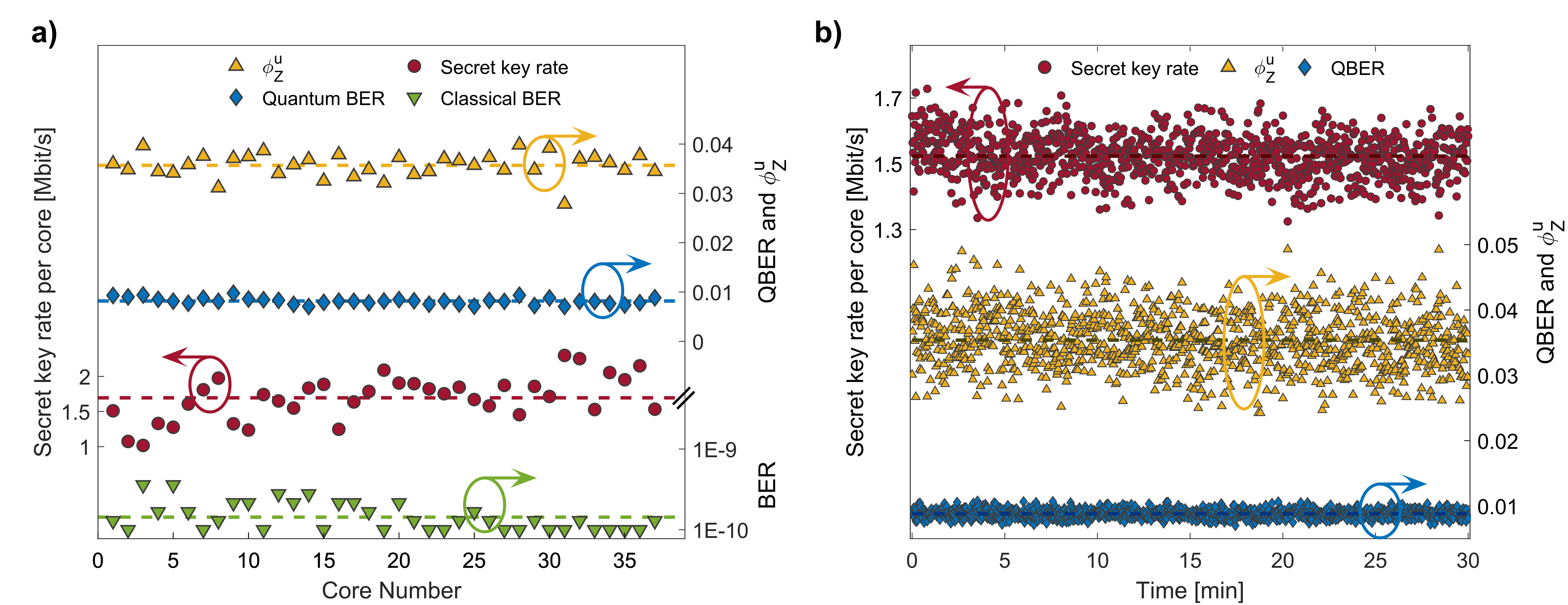}
\caption{\textbf{Quantum and classical transmission.} {\bf a) Secret key rate, bit error rate (BER), quantum bit error rate (QBER) and upper bound on the phase error ($\phi_Z^u$) per core with co-propagating classical signals.} Red dots represent the secret key rate per core. Blue diamonds and green downwards triangles are are the QBER and BER of the quantum and classical signals respectively. Yellow upwards triangles show the upper bound on the phase error. An average value of 1.7 Mbit s$^{-1}$ $\pm$ 2.72 kbit s$^{-1}$, 0.82\% $\pm$ 0.0028\% and 2.4E-10 was measured for the secret key rate, QBER and BER respectively, while the upper bounds on the phase error show an average of 3.57\% $\pm$ 0.015\%. Uncertainty values, computed as standard error of the mean, are not displayed as error bars are covered by the markers. {\bf b) Secret key rate, QBER and upper bound on the phase error as a function of time.} Red dots are the secret key rate measured in core number 37 for 30 minutes. Blue diamonds are the observed QBER in the same core and amount of time and yellow triangles the obtained upper bounds on the phase error. Average values of 1.52 Mbit s$^{-1}$ $\pm$ 1.95 kbit s$^{-1}$ with 0.88\% $\pm$ 0.0023\% QBER and 3.54\% $\pm$ 0.014\% $\phi_Z^u$ is achieved. A total amount of 60 Gbit can thus be created over the multicore fibre in 30 minutes of measurement.}
\label{fig:qac_results}
\end{figure*}

One of the main challenges of current quantum communications is related to its integration with the existing network infrastructures. Indeed, the majority of deployed quantum links are implemented with dark fibres~\cite{Qiu2017}. However, to make quantum communications cost-effective and to allow a faster deployment of this technology, the same infrastructure should be shared between classical and quantum signals.
To demonstrate the full compatibility of our QKD system, each of the 37 cores was shared between the quantum and a fast optical classical channel. 
The quantum states, at 1550 nm, are prepared by cascaded intensity and phase modulators as reported in Figure~\ref{fig:setup_simple} and previously explained. 
The classical signal is created by a continuous wave laser working at $1558$ nm (100GHz dense wavelength division multiplexing channel C23) and then modulated by an intensity modulator with a $10$ Gbit s$^{-1}$ on-off keying format. After the propagation through the 7.9-km MCF, a wavelength division multiplexing (WDM) filter separates the classical from the quantum signal. The quantum signal is measured with single photon detectors, whereas the classical light is received by a pre-amplified direct-detection receiver and analyzed by an error analyzer.
Further details of the experimental setup are reported in the Methods.
Figure~\ref{fig:qac_results} a) shows the measured secret key rate per core acquired during the simultaneous co-propagation of the classical channel at 10 Gbit s$^{-1}$ (with bit error rate BER $<$ 1E-9) in all the 37 cores. The average secret key rate per core is 1.7 Mbit s$^{-1}$ $\pm$ 2.72 kbit s$^{-1}$ yielding a total rate of 62.8 Mbit s$^{-1}$ $\pm$ 100 kbit s$^{-1}$. The block size used corresponds to 30 seconds acquisition in this configuration, i.e. 3.58 Gbit (a lower value than the one used in the only quantum scenario, due to the extra filter insertion loss at Bob's side).
Figure~\ref{fig:qac_results} a) also shows the measured bit error rate both for classical and quantum channels with an average BER of 2.4E-10 and QBER of 0.82\% $\pm$ 0.0028\%, respectively. Finally, the upper bounds derived for the phase error ($\phi_Z^u$) are reported, with an average of 3.57 $\pm$ 0.015\%. As a stability investigation, we measure the system performances (secret key rate, QBER and $\phi_Z^u$) on the core with the highest cross-talk (central core, number 37), with all the other cores lighted up, over a period of 30 minutes. The results are presented in Figure~\ref{fig:qac_results} b). The overall performance is stable during the long measurement period, with an average secret key rate of 1.52 Mbit s$^{-1}$ $\pm$ 1.95 kbit s$^{-1}$, an average QBER of 0.88\% $\pm$ 0.0023\% and an average $\phi_Z^u$ of 3.54\% $\pm$ 0.014\%.
The system is not actively stabilized and this performance can therefore be considered representative for future practical applications. During the 30 minutes measurement, a total key of approximately 60 Gbit is generated from the 37 cores, assuming similar statistics from all other cores. This value is lower than the expected secret key rate that could be obtained (i.e. $1.52 \text{ Mbit s$^{-1}$} \times 37 \times 30 \text{ min} \approx 101.2 \text{ Gbit}$); the only reason is that our real time QBER analysis is done by the same software interface that collects the detection events, and it therefore requires to shortly stop the photon acquisition. However, this could be easily improved with the use of two distinct software interfaces, which would lead to the expected secret key rate value.

\subsection*{Discussion}
MCFs represent the new frontier of optical communication, which in classical communications can be used for high capacity transmission and energy saving reasons. The major improvement over bundles of single mode fibres is given by space saving: standard fibers have one core in a 125 $\mu$m cladding diameter, while MCFs have several cores in a comparable cladding size. In applications where space is limited, such as data centres, the use of MCFs over bundles of single mode fibres is considered as a good candidate~\cite{Yuan2018}. Moreover, in less than 10 years of development the fabrication of MCFs has already reached performances (such as the loss in dB/km) comparable to those of standard single mode fibers~\cite{hayashi2012,hayashi2017}.
As a new application of MCFs for quantum communications, we devise a parallel quantum key distribution system that opens up the possibility of generating secret keys achieving the highest secret key generation rate of $105.7$ Mbit s$^{-1}$.
This result is achievable using a protocol that requires only few standard optics components, i.e. a simple and straightforward setup. It also requires the modulation and detection of only three states, instead of the usual four that constitute the two full mutually unbiased bases, which means that in principle only two single photon detectors are required. Moreover, it is known from literature that the modulation, stabilisation and detection of superposition states (those in the Fourier basis) are more challenging~\cite{Rusca2_2018,Ding2016,Cozzolino2018}: hence the use of this three-states protocol results in a further simplification of the system. Finally, our analysis is carried out in the finite key regime, with a block size corresponding to the actual sifted rate over 30 seconds (meaning that the time acquisition of our block is exactly 30 seconds). Hence, our performance metrics are close to a realistic implementation scenario. 
Concerning the co-propagation of quantum and classical channels, the reduction in secret key rate compared to the only quantum scenario is mainly caused by the insertion loss of the WDM filter at the output of each core. By adding approximately 3 dB of extra loss at receiver side, the photons arriving at Bob's detector and composing the raw rate are almost halved, which is then reflected by the final secret key rate. Indeed, from the simulations shown in Figure~\ref{fig:quantum_results} b), the expected key rate that can be extracted at 6.75 dB loss is 59.7 Mbit s$^{-1}$, certifying that the lower key rate is only due to a more lossy receiver and not from remaining leakage or corrupted channel performance due to the classical co-propagating signals.
To be noted that the protocol implemented in our work is a discrete variable (DV) modulation system. Usually quantum key distribution using squeezed states and homodyne detection, i.e. with a continuous variable (CV) modulation, is thought to be a promising candidate for practical quantum-cryptography implementations, due to its compatibility with existing telecom equipment and high detection efficiencies.
However, with this proof-of-concept experiment we prove that the co-existence of classical and quantum communications within the same spatial channels in a DV system is indeed feasible, showing that both DV and CV modulations can be considered telecom compatible. Actually, with an expected secret key generation rate of 62.8 Mbit s$^{-1}$, we are able to surpass any state of the art experiment (generating secret keys of at least one order of magnitude less), based either on DV \cite{Dynes2016_m,Dynes2016,Mao2018,Wang2017} or on CV QKD \cite{Huang2016,Eriksson2018_1,Eriksson2018_2,Eriksson2019}.
Furthermore, in its present configuration, the fibre length was limited to 7.9 km, but longer transmission distances are expected to be achievable as long as the power coming from the classical channel can be effectively filtered out while being sufficient for a low BER detection. In this work, the WDM filter used at the fibre output has a finite extinction ratio between channels of approximately 80 dB; thus, adjusting the input power of the classical channel such that the received power per core is around -34 dBm makes the power leakage unobservable on the quantum channel, \textit{i.e.} the QBER takes values similar to those it has during the quantum-only transmission (see Supplementary Note 3). Indeed, a received classical power of -34 dBm per core is enough for our setup to achieve 10 Gbit s$^{-1}$ communications with BER always lower than 10$^{-9}$; hence, in this configuration, transmitting at the full typical 0 dBm launch power is unnecessary for error-free detection.
As a further matter a real implementation of the system should be addressed: in particular, the ability to generate independent keys requires multiple intensity and phase modulators. In fact, a real deployment of this technology for quantum (and classical) communications requires independent data to be prepared and sent through the fibre.
This can be achieved using either conventional wavelength and space multiplexing technology in fibre or photonic integration, since several devices can be integrated onto the same chip. Fast and reliable photonic integrated circuits are already commercially available for classical communications, while quantum integrated photonics is a rapidly growing field~\cite{Ding2016,Sibson2015,Ma2016,sibson2017,Bunandar2018,semenenko2019,agnesi2019}.
At the same time, to achieve a very high rate for quantum communication, SNSPDs or fast gated InGaAs single photon detectors are needed. Some preliminary demonstrations guarantee the high compatibility of cryogenic temperature for optical receiver and photonic integrated circuits~\cite{Tyler2016,najafi2015}.
By combining integrated modulators and detectors on a photonic integrated circuit, an all-optical commercial system with high rate quantum and classical channels can be achieved. Finally, as a future perspective, to further boost the secret key rate a combined approach between SDM and WDM can be considered as already demonstrated for classical communications~\cite{Hu2018,Kobayashi2017,Puttnam2015}. 

In conclusion, we demonstrate through a proof-of-concept experiment that with SDM in a 37-core fibre it is possible to reach a quantum key generation rate of 105.7 Mbit s$^{-1}$ and, at the same time, the co-propagation of a reliable and error-free classical optical signal at 370 Gbit s$^{-1}$ is also feasible. By exploiting the hitherto most advanced multi-core fibre, we are able to overcome previous implementations paving the way for new quantum applications and for a faster deployment of quantum communications.  

\subsection*{Methods}
\subsubsection*{Experimental setup and electronic design}
Here we describe the setup reported in Supplementary Figure 1 and used to perform the experiment. To prepare a train of modulated weak coherent pulses, Alice (the transmitter) carves with two cascaded intensity modulators the time-encoded pulses from a tunable continuous wave laser, emitting at $1550$ nm (only one intensity modulator is reported in Supplementary Figure 1 for simplicity). The two cascaded intensity modulators are needed to enhance the extinction ratio of the weak coherent pulses. A third intensity modulator is used to implement the decoy state technique. The optical signal is then sent through a phase modulator which is used to phase randomize the quantum states. Subsequently, a variable optical attenuator is used to reach the quantum regime with a mean photon number of $\mu_1 \approx 0.11$ (for the signal intensity) and $\mu_2 \approx 0.07$ (for the decoy intensity) photon/pulse at the input of each core ($P_{QA}\approx$-81 dBm in Supplementary Figure 1). 
At Alice's side, four electrical outputs generated by a field programmable gate array are used to drive the three intensity modulators and for the synchronization signal. The phase modulator is driven by a digital-to-analog converter which uses 8 bit to obtain $2^{8}-1$ different phase values. The repetition frequency is $\nu = 595$ MHz, the electrical pulse width is approximately $100$ ps, whereas the obtained optical pulse width is around $150$ ps. 
For the classical signal, an intensity modulator allows on-off keying modulation at $10$ Gbit/s of light coming from a second laser emitting at $1558$ nm. The quantum and classical signals are combined with a beam combiner. Cascaded beam splitters (two 1$\times$5 and four 1$\times$8) are used to split the combined signal in 37 in order to couple them into the 37 cores. The fibre is a 7.9 km long heterogeneous MCF, \textit{i.e.}, it has three different types of cores with three different refractive indeces to decrease the inter-core cross-talks~\cite{Sasaki2017} (the measured cross-talk is shown in Supplementary Figure 2). At the receiver side (Bob's side), each core output is analysed consecutively (one by one): a wavelength division multiplexing filter separates the classical from the quantum signals (100GHz dense wavelength division multiplexing channel C23 and 100GHz dense wavelength division multiplexing channel C33, respectively). The classical signal, after being amplified and the out-of-band noise filtered out by an optical bandpass filter, is received by a photo detector and characterised by an error analyzer. The classical receiver performance in terms of BER for different optical received power values is shown in Supplementary Figure 3: with this receiver implementation a received optical power of -34 dBm is enough to obtain BER values always lower than 10$^{-9}$. Regarding the quantum channel, it is measured in the $\mathbf{X}$ and $\mathbf{Z}$ bases. A 10 dB beam splitter divides the two measurements. To maximize the key generation rate, the quantum signal in the $\mathbf{Z}$ basis is further divided by a beam splitter 1$\times$4 and connected to four different SNSPDs. This was necessary to avoid the detectors saturation regime, as it would lead to both a lower count rate and a higher QBER, due to the finite dead time of the devices and to the afterpulse effect respectively. In the $\mathbf{X}$ basis a single SNSPD, placed at one output of a delay line interferometer, is used to collect the photons. We implemented a free-space delay line interferometer, sitting on a damped optical table and enclosed in a box to increase stability and to minimize coupling of light from the room. A piezoelectric actuator on a mirror was used to finely tune the interference, in order to achieve high visibility. The visibility of the interferometer is shown in Supplementary Figure 4. All the electrical outputs are collected by a time tagging unit and analyzed by a computer. As a final comment, during the post-processing analysis the received clicks were temporally filtered: this was done to reduce the number of errors coming from detectors dark counts (and cross-talk in the case of co-propagating classical signals). The filters width in time was chosen such that the click rate after filtering was unchanged from the one obtained before filtering (the temporal pulse shape was all included in the filter) for all cases except from the point with largest attenuation, where the filters were chosen to be narrower to reduce even more the probability of dark counts and therefore increase the signal to noise ratio (see Supplementary Table 1 for the actual values used).

\subsubsection*{Data and code availability}
Data and code are available under reasonable requests.

\bibliographystyle{naturemag}
\bibliography{biblio}

\subsection*{End Notes}
\subsection*{Acknowledgments}
We thank D. Rusca for the fruitful discussion. Portions of this work were presented at the International Photonic Conference (IPC) in 2018~\cite{Dalio2018}.
The SPOC Centre for Silicon Photonics for Optical Communications (ref DNRF123); The bigQ Center for Macroscopic Quantum States (ref DNRF142); People Programme (Marie Curie Actions FP7$/$2007-2013- n 609405); VILLUM FOUNDATION Young Investigator Programme; the European Research Council through the ERC-CoG FRECOM project (grant agreement no. 771878); EU-Japan coordinated R\&D project SAFARI commissioned by the Ministry of Internal Affairs and Communications of Japan and EC Horizon 2020, QuantERA ERA-NET SQUARE project and by the International Network Programme SCQC.  

\subsection*{Author contributions}
D. B., B. D.~L., F. D.~R. and D. C. proposed the idea. D. B. and B. D.~L. carried out the theoretical analysis on the protocol. D. B., B. D.~L., D. C., F. D.~R. carried out the experimental work. D. B., B. D.~L., D. C., F. D.~R., X. G., Y. D., Y. S., K. A., S. M., H. T., T. Y., J. S. N.~N., M. G., K. R., U.L.~A., T. M. and L. K.~O discussed the results and contributed to the writing of the manuscript.

\subsection*{Competing financial interests}
The Authors declare no Competing Financial or Non-Financial Interests
\begin{table*}[h]
\begin{mdframed}[backgroundcolor=mycol,innerleftmargin=0pt,topline=false,rightline=false,leftline=false,bottomline=false]
    \centering
    \caption{\textbf{Comparison with state of the art experiments.} CC: classical channel; RX: receiver; QC: quantum channel; MUX: multiplexing; BB84: Bennet and Brassard 1984 protocol; SNSPD: superconductive nanowire single photon detector; CW: continuous wave; WDM: wavelength division multiplexing; SPD: single photon detector; ch.: channel; SDM: space division multiplexing.}
    \label{tab:comparison}
    \footnotesize
    \renewcommand{\arraystretch}{1.5}
    \begin{tabular}{p{1.4cm} p{1.45cm} p{1.5cm} p{1.4cm} p{1.4cm} p{1.2cm} p{1.45cm} p{1.6cm} p{1.8cm}}
         \toprule
          & \textbf{Rate CC,\newline Rx power} & \textbf{Rate QC} & \textbf{Distance,\newline loss} & \textbf{Protocol} & \textbf{Decoy} & \textbf{Operation} & \textbf{Classical MUX} & \textbf{Detector} \\ \hline
         This work (only QC) & - & 105.7 Mbit s$^{-1}$ & 7.9 km,\newline 3.75 dB & 3-states BB84 & 1-decoy & finite key & - & SNSPDs \\ 
         Ref.~\cite{Islame1701491} & - & 26 Mbit s$^{-1}$ & 4 dB & 4-D BB84 & 2-decoy & finite key & - & SNSPDs \\
         Ref.~\cite{Frohlich2017} & CW light,\newline -23 dBm & 10 kbit s$^{-1}$ & 100 km,\newline 18 dB & BB84 & 2-decoy & finite key & WDM\newline C-band & Self-differencing InGaAs SPDs \\
         Ref.~\cite{Dynes2016_m} & 20 Gbit s$^{-1}$,\newline -13.5 dBm per ch. & 605 kbit s$^{-1}$ & 53 km,\newline 13.5 dB & BB84 & 2-decoy & real-time & SDM+WDM C-band & Self-differencing InGaAs SPDs \\ 
         Ref.~\cite{Dynes2016} & 200 Gbit s$^{-1}$,\newline -36 dBm per ch. & 1.9 Mbit s$^{-1}$ & 35.5 km,\newline 6.8 dB & BB84 & 2-decoy & finite key & WDM C-band & Self-differencing InGaAs SPDs \\ 
         This work (QC+CC) & 370 Gbit s$^{-1}$,\newline -34 dBm per ch. & 62.8 Mbit s$^{-1}$ & 7.9 km,\newline 3.75 dB & 3-states BB84 & 1-decoy & finite key & SDM+WDM C-band & SNSPDs \\ 
         Ref.~\cite{Mao2018} & 3.6 Tbit s$^{-1}$,\newline 8.7 dBm & 4.5 kbit s$^{-1}$ & 66 km,\newline 19.83 dB & BB84 \newline(O-band) & 2-decoy & real-time & WDM C-band & Self-differencing InGaAs SPDs \\ 
         Ref.~\cite{Wang2017} & 6.38 Tbit s$^{-1}$,\newline -1 dBm & 14.8 kbit s$^{-1}$ & 50 km,\newline 16.5 dB & BB84 \newline(O-band) & 2-decoy & real-time & WDM C-band & Self-differencing InGaAs SPDs \\
         \bottomrule
    \end{tabular}
    \end{mdframed}
\end{table*}

\end{document}